\documentclass[reprint,amsmath,amssymb,aps,superscriptaddress,nofootinbib,floatfix]{revtex4-2}

\usepackage{graphicx}
\usepackage{dcolumn}
\usepackage{bm}
\usepackage{caption}
\usepackage{subcaption}
\usepackage{hyperref}
\usepackage{xcolor}
\usepackage[usestackEOL]{stackengine}

\begin{document}
\strutlongstacks{T}
\newcommand{\St}[1]{\Centerstack{#1}}

\title{Nature of the $P_c$ states from compositeness criteria}

\author{Yu-Fei Wang}
\email{wangyufei@ucas.ac.cn}
\affiliation{School of Nuclear Science and Technology, University of Chinese Academy of Sciences, Beijing 101408, China}

\author{Chao-Wei Shen}
\email{shencw@hdu.edu.cn}
\affiliation{School of Science, Hangzhou Dianzi University, Hangzhou 310018, China}

\author{Deborah R{\"o}nchen}
\email{d.roenchen@fz-juelich.de}
\affiliation{Institute for Advanced Simulation (IAS-4), Forschungszentrum J\"ulich, 52425 J\"ulich, Germany}

\author{Ulf-G.~Mei\ss ner}
\email{meissner@hiskp.uni-bonn.de}
\affiliation{Helmholtz-Institut f\"ur Strahlen- und Kernphysik (Theorie) and Bethe Center for Theoretical Physics, Universit\"at Bonn, 53115 Bonn, Germany}
\affiliation{Institute for Advanced Simulation (IAS-4), Forschungszentrum J\"ulich, 52425 J\"ulich, Germany}
%\affiliation{Tbilisi State University, 0186 Tbilisi, Georgia}
\affiliation{Peng Huanwu Collaborative Center for Research and Education, International Institute for Interdisciplinary and Frontiers, Beihang University, Beijing 100191, China}

\author{Bing-Song Zou} 
\email{zoubs@mail.tsinghua.edu.cn}
\affiliation{Department of Physics, Tsinghua University, Beijing 100084, China}
\affiliation{CAS Key Laboratory of Theoretical Physics, Institute of Theoretical Physics, Chinese Academy of Sciences, Beijing 100190, China}
%\affiliation{School of Physics, University of Chinese Academy of Sciences (UCAS), Beijing 100049, China}

\author{Fei Huang}
\email{huangfei@ucas.ac.cn (corresponding author)}
\affiliation{School of Nuclear Science and Technology, University of Chinese Academy of Sciences, Beijing 101408, China}

\begin{abstract}
Based on a coupled-channel approach, we investigate the structures of four $P_c$ states through compositeness criteria. Toward a more precise description of the states, we have obtained refined fit results of the LHCb data on the $J/\psi p$ invariant mass distribution of the $\Lambda_b^0\to J/\psi p K^-$ decay. Allowing for the fact that each of the four $P_c$ states couples strongly to a nearby $S$-wave channel, three criteria on the compositeness/elementariness are adopted in this study: the pole-counting rule, the spectral density function, and the Gamow wave function. Compositeness information is extracted from the scattering amplitudes and the pole parameters (pole positions and residues), without any preconceived assumptions on the nature of the $P_c$ states , and without any dependence on the model parametrization. Consistently within the framework of all the three methods, it has been found that the $P_c(4312)\,1/2^-$ is mainly composed by $\bar{D}\Sigma_c$,  $P_c(4380)\,3/2^-$ by $\bar{D}\Sigma_c^*$, while the $P_c(4440)\,1/2^-$ and $P_c(4457)\,3/2^-$ states both turn out as composite states of $\bar{D}^*\Sigma_c$. The upper limits of the values of their elementariness are estimated to be rather small. This paper provides an additional confirmation of the molecular interpretation for the $P_c$ states in the literature. 
\end{abstract}

\maketitle

\section{Introduction}
Since the discovery of the $X(3872)$ state in 2003~\cite{X3872}, the exotic hadron states have opened a new window for the understanding of how hadrons are formed. They indicate multiple mechanisms of the structures beyond the conventional picture in which a hadron is a compact cluster of a quark and an antiquark (mesons) or three quarks (baryons). This has intrigued the community to find new interpretations. The so-called  ``Hadronic molecule''~\cite{Guo:2017jvc} is actually a reasonable interpretation of many exotic states, describing such states as bound states of hadrons formed by residual interactions, given by hadron exchanges. Toward a deeper understanding of QCD, the examination of the ``molecular nature'' for the exotic states is a meaningful step. 

Specifically, as hidden-charm pentaquark candidates, the $P_c$ states are typically exotic and frequently discussed in the literature. In 2015, a very broad $P_c(4380)$ state with a narrow $P_c(4450)$ were observed for the first time by the LHCb collaboration in the $J/\psi p$ invariant mass distribution of the decay process $\Lambda_b\to J/\psi p K^-$~\cite{LHCb:2015yax}. Updated results from LHCb~\cite{LHCb:2019kea} showed that $P_c(4450)$ actually corresponds to two states $P_c(4440)$ and $P_c(4457)$, with an additional state $P_c(4312)$. The $P_c(4380)$ state has become insignificant yet not excluded in the updated analyses. There are also some indications for the existence of $P_c(4337)$~\cite{LHCb:2021chn}. Note that the masses, widths, and quantum numbers of such $P_c$ states still demand confirmation from theoretical studies as well as further experimental measurements by LHCb~\cite{Johnson:2024omq}. Interestingly, many theoretical studies~\cite{Wu:2010jy,Wu:2010vk,Wang:2011rga,Wu:2012md,Burns:2015dwa,Chen:2015loa,Chen:2015moa,He:2015cea,Meissner:2015mza,Chen:2016heh,Chen:2016qju,Lu:2016nnt,Ortega:2016syt,Roca:2016tdh,Shen:2016tzq,Lin:2017mtz,Du:2019pij,Guo:2019kdc,Liu:2019tjn,Liu:2019zvb,Meng:2019ilv,Xiao:2019aya,Xiao:2020frg,Yao:2020bxx,Du:2021fmf,Wang:2022ltr,Li:2025ejt} indicate or assume that the $P_c$ states are $S$-wave near-threshold hadronic molecules of $\bar{D}^{(*)}\Sigma_c^{(*)}$. Though at least for some of the $P_c$ states, there could be other possibilities such as kinematic cusps~\cite{Kuang:2020bnk,Nakamura:2021qvy,Burns:2022uiv} or compact pentaquark configurations~\cite{Ali:2019npk,Eides:2019tgv,Pimikov:2019dyr,Wang:2019got,Weng:2019ynv,Zhu:2019iwm,Giron:2021fnl,Ruangyoo:2021aoi,Santos:2024bqr}. 

The investigation of the structures of the $P_c$ states is usually tied up with the determination of the spectroscopy and the decay processes. As for the structures, discussions in the literature are mainly twofold: first, $P_c$ states are believed to be hadronic molecules due to the fact that they can be dynamically generated from the interactions of the components, and, second, the decay widths can be compatibly reproduced under the assumption that they are hadronic molecules. However, to gain deeper insights into their nature, it is better to carry out particular analyses on the compositeness {\it without}  prejudice. Especially, whether a state can be dynamically generated, in principle, cannot serve as a solid criterion on its structure. On the one hand, ``dynamical generation'' depends much on the technical details of the model involved in the study; on the other hand, elementary states can also be dynamically generated. A practical example is our recent study on the nature of the $N^*$ states~\cite{Wang:2023snv}: even if the $N^*(1440)$ resonance is dynamically generated in the model, it still gains unignorable elementariness: see Tables III and IV in Ref.~\cite{Wang:2023snv}, in which in all three criteria, the elementariness of $N^*(1440)$ is larger than $30\%$. The spectral density function of the model even gives a value about $50\%$.  Therefore we recommend systematic studies on the compositeness of the $P_c$ states. 

The origin of such a kind of studies is Weinberg's paper on the deuteron~\cite{weinberg1965}, the spirit of which is the connection between the probability of the state to be elementary (elementariness) and the near-threshold behavior of the amplitude. The latter can either be scattering length/effective range, low-energy constants, or the pole position and residue (coupling) of the state -- anyhow related to the experimental data and insensitive to the model artifacts~\cite{vanKolck:2022lqz}. Unlike the deuteron, exotic hadron states are resonances on which Weinberg's criterion cannot be applied directly. There are mainly three modified criteria for them. First is the pole-counting rule~\cite{Morgan:1992ge}, which infers the nature of a resonance by counting the number of the nearby poles. As a qualitative criterion, the pole-counting rule is valid only for the resonances coupling through $S$-wave to a nearby threshold, and similar to Weinberg's criterion, it does not depend on the details of certain models. The pole-counting rule has been applied on many hadron exotic states~\cite{Zhang:2009bv,Dai:2012pb,Meng:2014ota,Gong:2016hlt,Cao:2019wwt,Cao:2020gul,Kuang:2020bnk,Chen:2021tad,Santos:2024bqr}. Second, the spectral density function (SDF) method is the extension of Weinberg's elementariness as a distribution function on the energy region near the resonance peak. It was first proposed in Ref.~\cite{Baru:2003qq} and then applied on various hadrons~\cite{Kalashnikova:2005ui,Kalashnikova:2009gt,Baru:2010ww,Hanhart:2010wh,Hanhart:2011jz,Gong:2016hlt,Chen:2021tad,Wang:2023snv}. Mathematically, the SDF can be defined for any states, but to avoid model dependence, and to maintain its physical connection to Weinberg's criterion, the states still need to be close to the threshold of a dominate $S$-wave channel. The third criterion is based on the definition of the resonances as Gamow states~\cite{Gamow:1928aa,CIVITARESE200441} that are non-normalizable complex eigenstates of the Hamiltonian. For the mathematical descriptions of such states, see e.g. Refs.~\cite{ROMO1968617,Xiao:2016dsx,Xiao:2016wbs,Xiao:2016mon}. The quantities that naively correspond to Weinberg's compositeness can be calculated from the wave functions of the resonances~\cite{Sekihara:2016xnq}, being complex-valued. Transformations or extra measures should be employed to convert them to ``probabilities''. Applications of this kind of methods include but are not limited to Refs.~\cite{Aceti:2012dd,Hyodo:2013iga,Guo:2015daa,Meissner:2015mza,Sekihara:2015gvw,Sekihara:2016xnq,Oller:2017alp,Guo:2019kdc,Sekihara:2021eah,Matuschek:2020gqe,Wang:2023snv,Kinugawa:2024crb,Lin:2025pyk}. In fact, the applications of the compositeness criteria above on the $P_c$ states are rather sparse in the literature, though they are still mentioned; for example in Refs.~\cite{Kuang:2020bnk,Santos:2024bqr} the pole-counting rule is employed, while in Refs.~\cite{Meissner:2015mza,Guo:2019kdc} the complex compositeness $X$ is adjusted and the dependence of the decay widths is explored. 

In this paper, we carry out the analyses of the $P_c$ states based on our recent study in Ref.~\cite{Shen:2024nck}: the LHCb data has been analyzed by the J{\"u}lich-Bonn model. This model is a comprehensive dynamical coupled-channel (DCC) approach~\cite{Doring:2025sgb} for various meson-baryon scatterings. It originated in the studies of the pion-nucleon induced reactions and then extended to the hidden-charm reactions as well as the photoproduction reactions. For the details of this model, see Refs.~\cite{Ronchen:2022hqk,Wang:2022osj,Mai:2023cbp,Wang:2024byt,Wang:2022oof,Shen:2024nck,Doring:2025sgb} and the references therein. The channel space contains $J/\psi N$, $\bar{D}^{(*)}\Lambda_c$, and $\bar{D}^{(*)}\Sigma_c^{(*)}$, with the energy ranging from $4.3$ to $4.5$ GeV and the total angular momentum up to $7/2$. That study results in four narrow states with quantum numbers ($J^P$): $P_c(4312)\,1/2^-$, $P_c(4380)\,3/2^-$, $P_c(4440)\,1/2^-$, and $P_c(4457)\,3/2^-$. The resonance poles and their couplings (residues) to different channels are extracted in three fit solutions that all describe the data well. The comparisons thereof provide estimates of the systematical uncertainties. Note that a study of the compositeness of the $N^*$ and $\Delta$ resonances in the J{\"u}lich-Bonn model (light meson-baryon sector) has already been performed in Ref.~\cite{Wang:2023snv}. Owing to the fact that all the four $P_c$ states couple in $S$-wave to a nearby $\bar{D}^{(*)}\Sigma_c^{(*)}$ channel, in this paper, we can adopt all the three criteria and can also formulate the SDFs/Gamow wave functions without any extra parameters taken from the model (like the cut-off parameters). In other words, this study is independent of the parametrization of the model -- the model only provides the pole parameters as the inputs for those three criteria. Nevertheless, the residues of the poles may not be very stable in Ref.~\cite{Shen:2024nck} (for example, the complex coupling $g_{\bar{D}\Sigma_c}$ in Table 3 of Ref.~\cite{Shen:2024nck}), indicating remaining ambiguities. Hence, we perform refined fits in this study to get slightly better descriptions of the data and more precise pole parameters. The uncertainties from the pole parameters has indeed been decreased, as will be discussed in Sec.~\ref{sec:res}. 

This paper is organized as follows. In Sec.~\ref{sec:th} we briefly overview the theoretical foundations of the three compositeness criteria, and explain how they are related to the outputs of this coupled-channel approach. In Sec.~\ref{sec:res} we show the results of our refined fits, give the outputs of all the three criteria, and discuss relevant physics. Finally, Sec.~\ref{sec:con} is the conclusion. Two technical details which some readers may not be so interested in, namely the subtraction of the cusps in the SDFs and the phase convention of the residues, are explained in Appendixes~\ref{app:cusp} and \ref{app:phase}, respectively. 

\section{Theoretical framework}\label{sec:th}
\subsection{Pole-counting rule}
The first criterion, the pole-counting rule~\cite{Morgan:1992ge}, stems from the effective range expansion (ERE), 
\begin{equation}\label{Sera}
    M(p)\equiv p\cot\delta(p)=-\frac{1}{a}+\frac{r}{2}p^2+\mathcal{O}(p^4)\ ,
\end{equation}
where $p$ is the momentum, $\delta$ is the $S$-wave phase shift, $a$ is the scattering length, and $r$ is the effective range. We start with the ERE of a single-channel scattering amplitude, 
\begin{equation}
    T(p)=\frac{1}{M(p)-ip}\simeq\frac{1}{rp^2/2-ip-1/a}\ ,
\end{equation}
which contains two poles $p_\pm$ from the equation $rp^2/2-ip-1/a=0$: 
\begin{equation}
    p_\pm=\frac{1\pm\sqrt{1-2r/a}}{r}i\ .
\end{equation}
We take a common case as the example: $a>0$ and $r<0$. Note that the pole-counting rule also holds for other situations because the number of poles is the sole crucial factor. In this case, $p_-$ is a bound state (${\rm Im}p_->0$), and $p_+$ is a virtual state (${\rm Im}p_+<0$) lying farther away from the threshold ($|p_+|>|p_-|$). In the complex energy $E$ plane, $E_-$ is on the physical Riemann sheet, while $E_+$ is on the unphysical sheet. Hereafter, $p_-$ is called the ``main pole'' and $p_+$ is called the ``shadow pole''. Considering $p_++p_-=2i/r$, the poles are constrained by
\begin{equation}\label{polecons}
    |p_+|+|p_-|\geq \frac{2}{|r|}\ .
\end{equation}
Namely when the main pole is a bound state lying within the region $|p|\leq 1/|r|$, its shadow must lie beyond that region. For a typical quantum mechanical potential from hadron exchanges, the effective range $r$ is approximately equal to the interaction range $L$, which is at the fm level. Therefore, in the picture of typical potential scattering (molecular picture), within the range of one or two hundred ${\rm MeV}$s near the threshold, only the main pole can be found, while the shadow pole is absent. On the contrary, if the interaction is dominated by a Castillejo-Dalitz-Dyson (CDD) pole~\cite{Castillejo:1955ed} (a genuine state coupling weakly to the scattering channel), then the effective range $r$ is extremely large. Hence, Eq.~\eqref{polecons} loses its constraining power, and there can be two poles near the threshold. In fact, a CDD pole introduces two poles that are located almost equally close to the threshold~\cite{Morgan:1992ge}, i.e., $|p_+|\simeq|p_-|$. 

Therefore, practically, when one finds a single bound state near the threshold on the physical sheet, one can investigate its compositeness by looking for its virtual shadow pole. If there is a nearby shadow pole, one claims the bound state to be elementary; if there is none, it tends to be composite. The same rule is applicable for quasi-bound states (bound states with narrow widths) in coupled-channel scatterings. 

Considering the ERE in Eq.~\eqref{Sera}, the pole-counting rule is valid only for $S$-wave scatterings in the near-threshold energy region. In addition, the pole-counting rule is a qualitative method which does not provide a definite value for the ``elementariness'' or ``compositeness''. However, when it is valid, it is rather universal, regardless of the details of the models. 
\subsection{Spectral density functions}
The second (spectral density functions) and third (Gamow wave functions) criteria have already been explained in a detailed manner in Ref.~\cite{Wang:2023snv} with a solvable toy model. Here, we only repeat the central conceptions.  

The spectral density function~\cite{Baru:2003qq} (SDF) is a direct extension of Weinberg's elementariness. For a bound state $|B\rangle$, Weinberg's elementariness is defined as $Z\equiv 1-\int d\alpha|\langle\alpha|B\rangle|^2$, with $|\alpha\rangle$ being the two-body continuum of the corresponding configurations; for example, when studying the deuteron, $\alpha$ is the two-body continuum of a proton and a neutron. If the state is a narrow resonance with mass $M$ and width $\Gamma$, one substitutes $|B\rangle$ for a set of physical scattering states $|\Phi(z)\rangle$ with the energy $z$. Then, the elementariness $Z$ disperses into a finite distribution $w(z)$ which concentrates in the region $z\in [M-\Delta,M+\Delta]$, where the parameter $\Delta$ is comparable with the width $\Gamma$: usually chosen between $\Gamma/2$ and $2\Gamma$. The elementariness of such a resonance can be evaluated by collecting the SDF: $Z\simeq \int_{M-\Delta}^{M+\Delta} w(z) dz$. When only $S$-wave channels close to the mass are considered, $w(z)$ takes the non-relativistic Flatt{\'e} form~\cite{Kalashnikova:2009gt}, 
\begin{equation}\label{FSDF}
    w(z)=\frac{1}{2\pi}\frac{\sum_j g_j^2 p_j\Theta(z-\sigma_j)}{\big|z-M_0+\sum_n i g_n^2 p_n/2\big|^2}\ ,
\end{equation}
where $n$ and $j$ are the channel indices, $g_j$ is the real coupling constant for channel $j$, $p_j$ is the corresponding on-shell momentum, $\sigma_j$ represents the threshold of channel $j$, $M_0$ is a real mass parameter, and $\Theta$ is the Heaviside step function. Note that when the energy $z$ is below the threshold of channel $n$, $p_n$ is purely imaginary and the $i g_n^2 p_n/2$ term in the denominator is real-valued. 

The SDF has two important properties. First, it must be positive-definite, which is guaranteed when the couplings ($g_j$'s) in Eq.~\eqref{FSDF} are real. Second, without other bound states, the SDF must obey the sum rule ($\sigma_1$ is the lowest threshold): 
\begin{equation}\label{SR}
    \int_{\sigma_1}^{+\infty}dz\,w(z)=1\ .
\end{equation}
This is also guaranteed by the expression Eq.~\eqref{FSDF}. If a channel $n$ couples to the resonance with the orbital angular momentum $L\neq 0$, then the corresponding term in Eq.~\eqref{FSDF} is $g_n^2 p_n^{2L+1}$ rather than $g_n^2 p_n$. In this case one needs an extra regulator $F_n$ (containing extra cut-off parameters), i.e. $g_n^2 p_n^{2L+1}\to g_n^2 p_n^{2L+1}F_n$; otherwise, the sum rule~\eqref{SR} is violated. 

In practice, to suppress the ambiguity arising from the choice of the aforementioned $\Delta$ parameter, we take $\Delta=\Gamma$ and simultaneously estimate the elementariness with the help of the Breit-Wigner SDF: 
\begin{equation}\label{SDFZ}
\begin{split}
    BW(z)&\equiv\frac{1}{\pi}\frac{\Gamma/2}{(z-M)^2+(\Gamma/2)^2}\ ,\\
    Z&\simeq\frac{\int_{M-\Gamma}^{M+\Gamma}w(z)dz}{\int_{M-\Gamma}^{M+\Gamma}BW(z)dz}\ .
\end{split}
\end{equation}
This formula is also adopted in Ref.~\cite{Wang:2023snv}. Note that sometimes a higher threshold lies within the integral interval $[M-\Gamma,M+\Gamma]$, leading to a cusp in $w(z)$ that is believed to be an extra contribution and needs to be subtracted~\cite{Kalashnikova:2009gt}. Though not affecting the results much here, in this work, we also perform this subtraction, with more detailed explanations given in Appendix~\ref{app:cusp}. 
\subsection{Gamow wave functions}
The complex compositeness of a Gamow state is also a direct extension of Weinberg's elementariness. Different from the philosophy of the aforementioned SDFs, the bound state $|B\rangle$ in Weinberg's criterion is substituted by Gamow states~\cite{Gamow:1928aa,CIVITARESE200441} $|R)$ and $|R^*)$ (here, ``$*$'' denotes complex conjugation), which are complex eigenstates of the Hamiltonian: 
\begin{equation}
    \hat{H}|R^{(*)})=\left(M\mp\frac{\Gamma}{2}i\right)|R^{(*)})\ ,\quad (R^*|R)=1\ .
\end{equation}
Note that $|R)$ and $|R^*)$ are not normalized. As required by Schwartz's reflection principle $f(R^*)=f^*(R)$, for the continuum $|\alpha\rangle$, $(R^*|\alpha\rangle=\langle\alpha|R^*)^*=\langle\alpha|R)$, so the compositeness and elementariness for a Gamow state can be written as 
\begin{equation}
    \beta(\alpha)\equiv \langle\alpha|R)\ ,\quad X\equiv \int_C d\alpha\, \beta^2(\alpha)\ ,\quad Z\equiv1-X\ ,
\end{equation}
where $\beta(\alpha)$ is called ``Gamow wave function'', and the contour $C$ is $[0,\infty)$ plus a residue term at the resonance pole when the resonance is not on the physical Riemann sheet (see Ref.~\cite{Wang:2023snv} for details). In fact, the wave function can be obtained from the off-shell residues of the scattering amplitude at the resonance pole~\cite{Sekihara:2016xnq}, 
\begin{equation}
\begin{split}
    \langle\alpha'|\hat{T}|\alpha\rangle&\sim \frac{\gamma(\alpha')\gamma(\alpha)}{z-z_p}\ ,\\
    \beta(\alpha)&=\gamma(\alpha)G(z_p,\alpha)\ ,
\end{split}
\end{equation}
where $z_p=M-i\Gamma/2$ is the pole position and $G$ is the two-body propagator. The quantities $Z$ and $X$ in this regime are not real. To make comparisons with the second criterion, one should perform mathematical transformations or redefinitions so that the new quantities are real numbers between $0$ and $1$. For a ``quasi-bound state'' which couples in $S$-wave to a nearby channel, the imaginary parts of $X$ and $Z$ should be automatically small, while the real parts are close to Weinberg's case. 
\subsection{Coupled-channel approach}
\begin{figure*}[t!]
	\centering
	\includegraphics[width=0.6\textwidth]{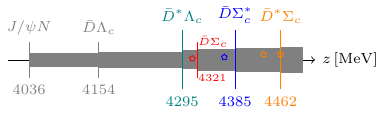}
	\caption{The relevant scattering channels and the thresholds. The four pentagrams label the four $P_c$ poles from the fit A of Ref.~\cite{Shen:2024nck}: from left to right, they are $P_c(4312)\,1/2^-$, $P_c(4380)\,3/2^-$, $P_c(4440)\,1/2^-$, and $P_c(4457)\,3/2^-$, respectively. }
	\label{fig:thr}
\end{figure*}

In this study, the pole positions and residues for the compositeness criteria are provided by the J{\"u}lich-Bonn model, based on the Lippmann-Schwinger-like equation
\begin{equation}\label{Tequ}
    \begin{split}
        &T_{\mu\nu}(p'',p',z)\\
        &=V_{\mu\nu}(p'',p',z)\\
        &+\sum_{\kappa}\int_0^\infty p^2 dp V_{\mu\kappa}(p'',p,z)G_{\kappa}(p,z)T_{\kappa\nu}(p,p',z)\ ,
    \end{split}
\end{equation}
where $p'$ and $p''$ are the center-of-mass momenta of the initial and final state, respectively; $z$ is the center-of-mass energy, $V$ is the interaction potential, $T$ is the scattering amplitude, and $G$ is the propagator of the intermediate channel. The greek letters $\mu$, $\nu$, and $\kappa$ are channel indices: for $J/\psi N$, $\bar{D}^{(*)}\Lambda_c$, and $\bar{D}^{(*)}\Sigma_c^{(*)}$ states with total isospin $I=1/2$ and with definite $JLS$ quantum numbers. The thresholds of the channels are shown in Fig.~\ref{fig:thr}. 
The model is constructed from time-ordered perturbation theory (TOPT)~\cite{schweber1964}, so the propagator takes the simple form
\begin{equation}
    G_{\kappa}(p,z)=\frac{1}{z-E_\kappa(p)-\omega_\kappa(p)+i0^+}\ ,
\end{equation}
where $E_\kappa(p)$, $\omega_\kappa(p)$ denote the energy of the baryon and meson in channel $\kappa$, respectively, in relativistic form: $E(p)=\sqrt{p^2+m^2}$. Moreover, the potential $V$ contains only $t$- and $u$-channel exchange diagrams, no $s$-channel resonance terms are included. In Ref.~\cite{Shen:2024nck} the $\Lambda_b\to K^- J/\psi p$ decay process has been calculated, with the amplitude $T$ in Eq.~\eqref{Tequ} describing the $J/\psi p$ final-state interaction. The model parameters, i.e., the couplings of $\Lambda_b K^-$ to the reaction channels, the coefficients of the polynomial for the background, and the cut-off parameters in the potentials,  have been determined through the fits to the LHCb data, resulting in three solutions that describe the data well. The pole positions of the four $P_c$ states and their couplings (residues) to the channels have also been worked out. 

The three criteria introduced in the last section are closely related to the output of this approach. Fortunately, unlike the study in Ref.~\cite{Wang:2023snv}, here the $P_c$ states are located in a rather narrow energy region, from the $\bar{D}^*\Lambda_c$ threshold ($4295$ MeV) to $\bar{D}^*\Sigma_c$ ($4462$ MeV), and each of them couples strongly to a nearby channel in the $S$ wave. Therefore the pole-counting rule is applicable, and it is reasonable to formulate the SDFs and the Gamow wave functions in a non-relativistic way without extra regulators (cut-off parameters). We consider thus only the $S$-wave channels in Table \ref{tab:SWch}. 
\begin{table}[t!]
    \small
    \renewcommand{\arraystretch}{1.2}
    \begin{ruledtabular}
    \begin{tabular}{ccc}
    States & $J^P$ & Channels in $S$-Wave\\
    \hline
    $P_c(4312)$ & $1/2^-$ & $\bar{D}\Sigma_c$, $\bar{D}^*\Lambda_c(a)$, $\bar{D}^*\Sigma_c(a)$\\ 
    $P_c(4380)$ & $3/2^-$ & $\bar{D}^*\Lambda_c(c)$, $\bar{D}^*\Sigma_c(c)$, $\bar{D}\Sigma_c^*(c)$\\ 
    $P_c(4440)$ & $1/2^-$ & $\bar{D}\Sigma_c$, $\bar{D}^*\Lambda_c(a)$, $\bar{D}^*\Sigma_c(a)$\\ 
    $P_c(4457)$ & $3/2^-$ & $\bar{D}^*\Lambda_c(c)$, $\bar{D}^*\Sigma_c(c)$, $\bar{D}\Sigma_c^*(c)$\\ 
    \end{tabular}
    \end{ruledtabular}
    \caption{Summary of the nearby channels coupling to the $P_c$ states in the $S$ wave. The combinations of the orbital angular momentum ($L$) and spin ($S$) are abbreviated as (a) $S=1/2$, (b) $S=3/2$, $|J-L|=1/2$, and (c) $S=3/2$, $|J-L|=3/2$. }
    \label{tab:SWch}
\end{table}

Specifically, we evaluate the compositeness in the following ways. 
\subsubsection{Pole-counting rule}
We perform analytical continuations on the scattering amplitude to the unphysical Riemann sheets corresponding to the virtual shadow poles of the $P_c$ states and then search for the virtual poles. This is done by deforming the contour in Eq.~\eqref{Tequ}, as discussed in Ref.~\cite{Doring:2009yv}. 
\subsubsection{SDFs}
Based on the general form in Eq.~\eqref{FSDF}, we locally construct the following expression for each $P_c$ state near its pole: 
\begin{equation}\label{SDF}
    w(z)=\frac{1}{2\pi}\frac{2h_0p_0\Theta_0+\sum_\kappa g_\kappa^2 p_\kappa\Theta_\kappa}
    {\big|z-M_0+\sum_\nu \frac{i}{2} g_\nu^2 p_\nu+ih_0p_0\big|^2}\ ,
\end{equation}
where $\kappa$ and $\nu$ are the channels in $S$ wave, as listed in Table \ref{tab:SWch}, and the momenta $p_\kappa=\sqrt{2\mu_\kappa(z-\sigma_\kappa)}$ are non-relativistic, with $\mu_\kappa$ being the reduced mass; the step functions $\Theta(z-\sigma_\kappa)$ are abbreviated as $\Theta_\kappa$. The couplings $g_\kappa$ are taken as the moduli of the normalized residues defined in PDG~\cite{PDG:2024cfk}
\begin{equation}\label{NRdef}
    g_\kappa=\Bigg|\sqrt{\frac{2\pi \rho_\kappa}{\Gamma}}r_\kappa\Bigg|\ ,
\end{equation}
where $r_\kappa$ is the residue of the $T$ amplitude in Eq.~\eqref{Tequ} and $\rho_\kappa=p_\kappa E_\kappa\omega_\kappa/z$ is a kinematic factor. As commonly assumed, the normalized residues measure the coupling strengths of a state to the scattering channels. Note that both the $g_\kappa$'s in Eq.~\eqref{SDF} and the normalized residues are dimensionless. The term with $h_0$ is the ``background contribution'', which embraces all the implicit contributions from the channels beyond Table \ref{tab:SWch}, and in the meantime guarantees that the starting point of the SDF is the $J/\psi N$ threshold and the pole position is correct. Specifically, $p_0=\sqrt{2\mu_0(z-\sigma_0)}$ corresponds to the lowest $J/\psi N$ channel. The parameter $h_0$ is not determined by the normalized residues of $J/\psi N$; together with $M_0$, it reproduces the pole position in the fit result. 

In a few words, Eq.~\eqref{SDF} is constructed only from the residues and pole positions determined by the fits. As mentioned in Ref.~\cite{Wang:2023snv}, such kind of construction may sometimes fail, violating the positive definite condition, i.e., $h_0<0$, which indicates uncertainties. Fortunately, in this study, the positive definiteness condition always holds.  
\subsubsection{Gamow wave functions}
We directly use the off-shell residues (the $\gamma$'s) of the $T$ amplitude from the fit results: 
\begin{equation}\label{offR}
    T_{\mu\nu}(p'',p',z)=\frac{\gamma_{\mu}(p'')\gamma_{\nu}(p')}{z-z_p}+\cdots\ .
\end{equation}
This definition, however, implies the symmetry condition $T_{\mu\nu}(p'',p',z)=T_{\nu\mu}(p',p'',z)$ which sometimes is not directly satisfied due to an extra phase $(-1)$. To maintain the consistency, the phases of the initial and final channels should be adjusted. For details, see Appendix~\ref{app:phase}. Similar to the formalism in Ref.~\cite{Wang:2023snv}, the complex compositeness of channel $\kappa$ and the elementariness are
\begin{equation}\label{XX}
\begin{split}
    X_\kappa&=\int_C p^2dp \frac{\gamma_{\kappa}^2(p)}{\big[z_p-\sigma_\kappa-p^2/(2\mu_\kappa)\big]^2}\ ,\\
    Z_{\rm max}&=1-\sum_\kappa X_\kappa\ ,
\end{split}
\end{equation}
where $\kappa$ denotes any channel in Table \ref{tab:SWch}. The contour $C$ is $[0,+\infty)$ when $z_p$ is on the physical sheet of channel $\kappa$; otherwise, $C$ also contains a residue term. Here, we also take the following measure~\cite{Sekihara:2021eah,Wang:2023snv} to convert the quantities in Eq.~\eqref{XX} to ``probabilities'': 
\begin{equation}\label{ZXp}
\begin{split}
    \tilde{X}_\alpha&=\frac{|X_\alpha|}{\sum_\kappa |X_\kappa|+|Z_{\rm max}|}\ ,\\
    \tilde{Z}_{\rm max}&=\frac{|Z_{\rm max}|}{\sum_\kappa |X_\kappa|+|Z_{\rm max}|}\ .
\end{split}
\end{equation}
\subsubsection{Remarks}
In this study, unlike Ref.~\cite{Wang:2023snv}, the $s$-channel genuine states are absent. Superficially, one might expect that the elementariness is trivially zero. However, this is not true. Physically, we have already explained that ``dynamical generation'' is not a sufficient condition to the composite nature. Note again that the compositeness criteria adopted here rely solely on pole parameters as inputs and are independent of the model's parametrization. Therefore, a model without $s$-channel poles would still yield nonzero elementariness. Technically, as for the pole-counting rule, it is of course possible for such complicated coupled-channel dynamics to generate a shadow pole near the quasi-bound state, leading to the elementary picture. The SDFs here are constructed from the pole parameters irrespective of whether the model has $s$-channel states or not, and indeed, they are non-zero. For the Gamow states, we only consider the $S$-wave channels and exclude the distant channels $J/\psi N$ and $\bar{D}\Lambda_c$, see Table \ref{tab:SWch}. Hence, the $Z_{\rm max}$ in Eq.~\eqref{ZXp} actually contains the compositeness from the other channels. Granted that the actual elementariness were zero, the $Z_{\rm max}$ would still be non-zero. That is why, instead of $Z$, we denote the result as $Z_{\rm max}$, the ``upper limit'' of the elementariness. In fact the actual elementariness cannot be zero either. Even in the wildest scenario with all the channels and partial waves included, the third criterion still allows non-zero elementariness when $s$-channel states are absent. See the discussions in Ref.~\cite{Sekihara:2016xnq}, as long as the potential explicitly depends on the center-of-mass energy $z$ (the ``missing channel contributions'' exist), the elementariness $Z$ can be non-zero. In a few words, the $Z_{\rm max}$ here is the sum of the compositeness values from the other channels and the intrinsic elementariness $Z$ from the missing channel contributions. 
\section{Results and discussions}\label{sec:res}
\subsection{Refined fits}\label{sec:refit}
Pole parameters (pole positions and residues) are crucial for the determination of the compositeness. In the previous study~\cite{Shen:2024nck}, the model has been used to fit the LHCb data~\cite{LHCb:2019kea}, and three fit solutions have been found. Though the three solutions describe the data almost equally well, the residues are not so stable, for example, the residue of $P_c(4312)$ in the $\bar{D}\Sigma_c$ channel in fit~B differs much to those in fits~A and C (see Table 3 of Ref.~\cite{Shen:2024nck}), and the situation is similar for the residues of $P_c(4380)$ in the $\bar{D}\Sigma_c^*$ channel. 

\begin{figure}[t]
	\centering
	\includegraphics[width=0.48\textwidth]{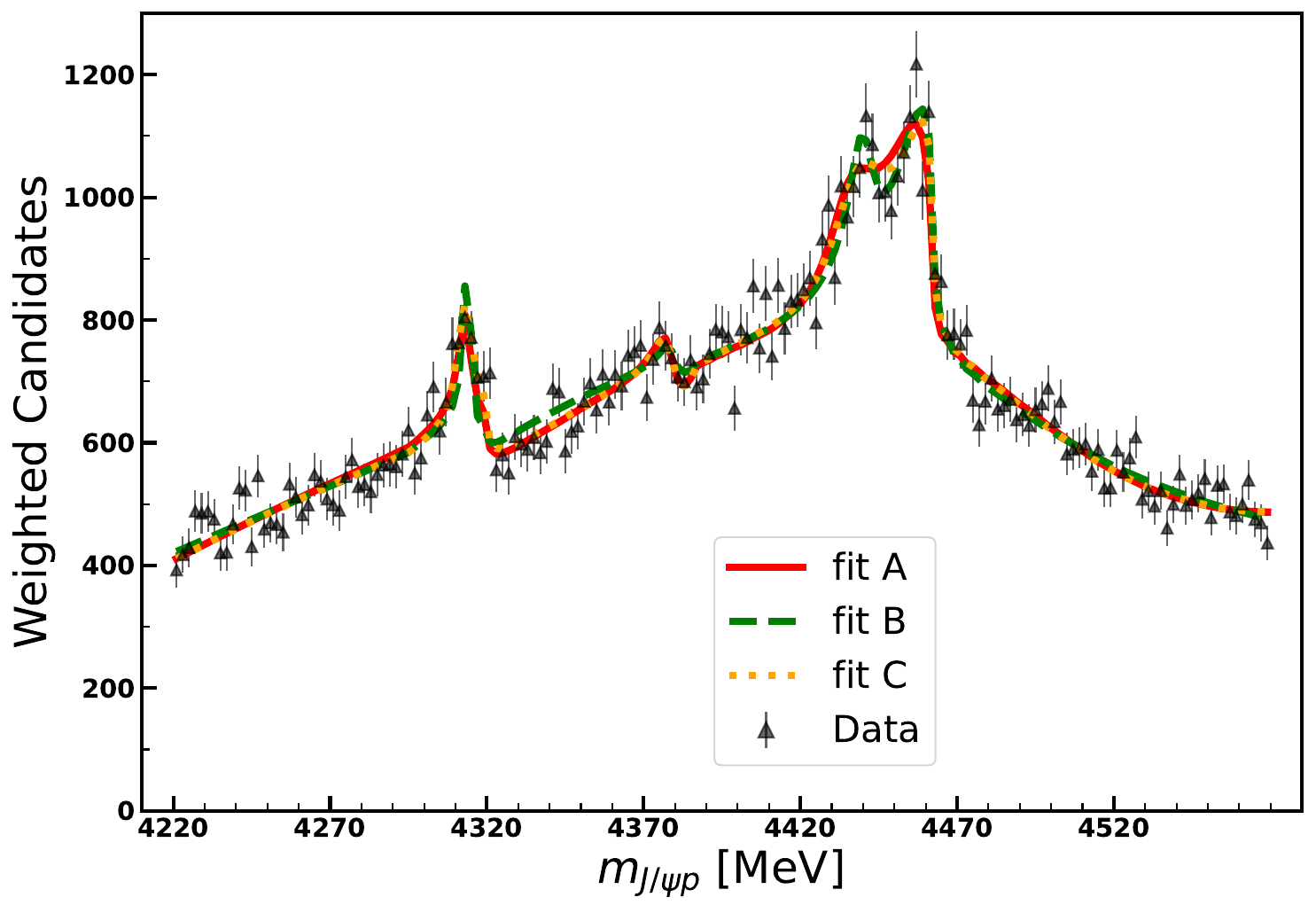}
	\caption{The three refined fit solutions in this study. The data are from Ref.~\cite{LHCb:2019kea}. }
	\label{fig:newfit}
\end{figure}

To gain a more precise description of the states, we have tried to perform refined fits to the data. Starting from the parameter values of the three fits in Ref.~\cite{Shen:2024nck}, we have manually added weights to the data points near the four $P_c$ states so that the parameter values can get out of the previous local minima of the $\chi^2$. Then, we withdraw the weights and continue the fits. In the end, we have obtained three new solutions with slightly better fit qualities, as depicted in Fig.~\ref{fig:newfit}. Here, we still name them as fits~A, B, and C according to which fit solution in Ref.~\cite{Shen:2024nck} they take as the starting values. It can be seen that compared to the previous results the double-peak structure around $4450$ MeV has been fitted better in fit~B. Note that the refits here are purely numerical: we have not changed any theoretical formulation of the model, the details of which can be found in Refs.~\cite{Wang:2022oof,Shen:2024nck}. Therefore, the set of fit parameters and the data are also exactly the same as in Ref.~\cite{Shen:2024nck}: $175$ data points and $76$ free parameters. 

The pole positions of the four $P_c$ states in the new solutions do not deviate very much from the old results. As shown in Table \ref{tab:Pcpoles}, they have the same quantum numbers as in Ref.~\cite{Shen:2024nck}, and they are all narrow. The masses of $P_c(4380)$ in fits A and C have been shifted to around $4378$ MeV in this study. It is worth mentioning that in the study by LHCb~\cite{LHCb:2019kea} the $P_c(4380)$ state is insignificant, yet still cannot be excluded. There are theoretical studies in the literature that also result in such a state; see, e.g., Refs.~\cite{Du:2019pij,Xiao:2020frg,Du:2021fmf,Burns:2022uiv}. Our model actually favors a relatively narrow $P_c(4380)$, since it behaves rather stably against different fits. Note that in all the three solutions here, $P_c(4457)$ lies a bit above the $\bar{D}^*\Sigma_c$ threshold ($4462$~MeV). This does not cause any physical problem since it is on the correct Riemann sheet for a quasi-bound state of $\bar{D}^*\Sigma_c$. It could usually happen that a bound state is pushed to be a little above the threshold when the lower channels open and complicated coupled-channel dynamics take effect. This phenomenon has already been seen in the fit~C of Ref.~\cite{Shen:2024nck}. In addition, there are some other distant poles in the model, cf. Table 4 of Ref.~\cite{Shen:2024nck}. Here we skip the discussions on them since i) the current data are not sufficient to firmly support their existence and ii) they are not related to the topic of this paper.  
\begin{table}[t!]
    \small
    \begin{ruledtabular}
    \begin{tabular}{cccc}
    States & Fit A & Fit B & Fit C\\
    \hline
    $P_c(4312)\, 1/2^-$ & $4313.4-2.2i$ & $4314.0-1.1i$ & $4312.6-2.9i$\\ 
    $P_c(4440)\, 1/2^-$ & $4435.7-9.2i$ & $4440.3-4.7i$ & $4438.3-9.9i$\\ 
    $P_c(4380)\, 3/2^-$ & $4378.2-4.5i$ & $4378.8-3.8i$ & $4378.1-4.3i$\\ 
    $P_c(4457)\, 3/2^-$ & $4463.6-12i$  & $4465.4-11i$  & $4466.3-9.0i$\\ 
    \end{tabular}
    \end{ruledtabular}
    \caption{The pole positions of the four $P_c$ states in the refined fit solutions of this study (in units of MeV). }
    \label{tab:Pcpoles}
\end{table}

For the residues (couplings) $r_\kappa$ in each channel $\kappa$, the definition is
\begin{equation}
    T_{\mu\nu}\simeq \frac{r_{\mu}r_{\nu}}{z-z_p}+\cdots\ .
\end{equation}
where $T$ is the scattering amplitude in Eq.~\eqref{Tequ} (set to be on-shell). As already mentioned, the factorized definition of the residues needs the symmetry condition $T_{\mu\nu}=T_{\nu\mu}$; therefore, we have adjusted the phases, as discussed in Appendix~\ref{app:phase}. In Table \ref{tab:res} we show those residues extracted from the three refined fits. The previous instabilities of the residues have been overcome. The following analyses of the compositeness are all based on the three refined solutions. 
\begin{table*}[t!]
%\small
\begin{ruledtabular}
	\begin{tabular}{ccccc} 
		Channels & $P_c(4312)\, 1/2^-$ & $P_c(4440)\, 1/2^-$ & $P_c(4380)\, 3/2^-$ & $P_c(4457)\, 3/2^-$ \\
		\hline
		$\bar{D}\Lambda_c$ & \St{$-1.1+0.082i$\\$-0.53+0.46i$\\$-1.2-0.14i$} & \St{$-1.7+0.30i$\\$-0.86+0.80i$\\$-2.0+0.095i$} & 
        \St{$0.19-0.0054i$\\$0.29-0.025i$\\$0.22+0.0017i$} & \St{$0.89+0.12i$\\$1.0+0.020i$\\$0.90+0.24i$}\\ 
        \hline
		$\bar{D}\Sigma_c$ & \St{$7.6+0.97i$\\$7.1+0.57i$\\$7.8+1.0i$} & \St{$-0.62-0.034i$\\$-0.062+0.64i$\\$-0.78-0.033i$} & 
        \St{$-0.017-0.012i$\\$-0.014-0.0073i$\\$(-9.0-8.8i)\times 10^{-3}$} & \St{$0.64+0.20i$\\$0.60+0.21i$\\$0.62+0.27i$}\\
        \hline
		$\bar{D}^*\Lambda_c\,(a)$ & \St{$0.39-0.10i$\\$0.40-0.13i$\\$0.37-0.10i$} & \St{$0.48-0.27i$\\$0.25-0.22i$\\$0.39-0.24i$} & 
        \St{$-0.72+0.0088i$\\$-0.76+0.012i$\\$-0.70+0.0054i$} & \St{$0.49+0.23i$\\$0.52+0.21i$\\$0.44+0.26i$}\\
        \hline
		$\bar{D}^*\Lambda_c\,(b)$ & $0$ & $0$ & \St{$-0.66-0.012i$\\$-0.69-0.013i$\\$-0.65-0.012i$} & \St{$0.66+0.18i$\\$0.65+0.19i$\\$0.53+0.19i$}\\
        \hline
		$\bar{D}^*\Lambda_c\,(c)$ & \St{$0.22+0.016i$\\$0.20-0.0024i$\\$0.23+0.00087i$} & \St{$-0.046-0.010i$\\$-0.084-0.055i$\\$-0.000046-0.013i$} & 
        \St{$1.1-0.64i$\\$0.89-0.51i$\\$1.1-0.62i$} & \St{$0.23+0.39i$\\$0.022+0.28i$\\$0.18+0.39i$}\\
        \hline
		$\bar{D}^*\Sigma_c\,(a)$ & \St{$6.9+1.2i$\\$7.3+0.68i$\\$6.8+1.5i$} & \St{$13+1.6i$\\$11+0.67i$\\$13+1.4i$} & 
        \St{$1.8+0.65i$\\$1.9+0.43i$\\$1.9+0.59i$} & \St{$0.044+0.019i$\\$0.050+0.023i$\\$0.039+0.047i$}\\
        \hline
		$\bar{D}^*\Sigma_c\,(b)$ & $0$ & $0$ & \St{$4.9-1.5i$\\$4.7-1.4i$\\$4.9-1.5i$} & \St{$-0.070-0.20i$\\$-0.046-0.19i$\\$-0.014-0.19i$}\\
        \hline
		$\bar{D}^*\Sigma_c\,(c)$ & \St{$12-1.8i$\\$9.6-1.4i$\\$12-2.1i$} & \St{$-0.60-0.23i$\\$-0.47-0.20i$\\$-0.61-0.24i$} & 
        \St{$-4.9-0.18i$\\$-4.4+0.10i$\\$-4.5-0.15i$} & \St{$-6.3-5.0i$\\$-5.9-5.1i$\\$-5.4-4.9i$}\\
        \hline
		$\bar{D}\Sigma_c^*\,(b)$ & $0$ & $0$ & \St{$0.098+0.023i$\\$0.092+0.031i$\\$0.092+0.023i$} & \St{$-0.085-0.066i$\\$-0.045-0.094i$\\$-0.042-0.092i$}\\
        \hline
		$\bar{D}\Sigma_c^*\,(c)$ & \St{$0.052-0.41i$\\$-0.0033-0.48i$\\$0.11-0.39i$} & \St{$-0.0068-0.091i$\\$0.059-0.043i$\\$-0.0099-0.17i$} & 
        \St{$7.1+2.1i$\\$6.9+1.9i$\\$7.2+2.0i$} & \St{$0.43+2.8i$\\$0.46+2.8i$\\$0.26+2.7i$}\\
        \hline
		$J/\psi N\,(a)$ & \St{$0.0066+0.020i$\\$0.022+0.0060i$\\$0.0061+0.025i$} & \St{$0.22+0.020i$\\$0.22+0.014i$\\$0.25+0.024i$} & 
        \St{$(3.6+0.94i)\times 10^{-3}$\\$(2.7+0.63i)\times 10^{-3}$\\$(3.5+0.90i)\times 10^{-3}$} & \St{$(-7.4-4.8i)\times 10^{-3}$\\$(-7.2-4.3i)\times 10^{-3}$\\$(-6.9-5.4i)\times 10^{-3}$}\\
        \hline
		$J/\psi N\,(b)$ & $0$ & $0$ & 
        \St{$(-9.2-3.8i)\times 10^{-3}$\\$(-7.0-2.7i)\times 10^{-3}$\\$(-8.7-3.6i)\times 10^{-3}$} & \St{$0.021+0.012i$\\$0.021+0.011i$\\$0.019+0.013i$}\\
        \hline
		$J/\psi N\,(c)$ & \St{$(1.7+0.63i)\times 10^{-3}$\\$(3.2-0.074i)\times 10^{-3}$\\$(1.8+1.1i)\times 10^{-3}$} & \St{$0.042-0.0029i$\\$0.041-0.00071i$\\$0.048-0.0032i$} & 
        \St{$-0.16-0.064i$\\$-0.12-0.047i$\\$-0.15-0.061i$} & \St{$0.29+0.16i$\\$0.28+0.15i$\\$0.26+0.17i$}\\
    \end{tabular}
\end{ruledtabular}
\caption{The residues of the $P_c$ states in each channel (in units of $10^{-3}$ MeV$^{-1/2}$). The notations of the subchannels (a), (b) and (c) are the same as in Table \ref{tab:SWch}. In each cell, the three values from the top to the bottom represent the residues in the refined fits~A, B, and C, respectively. }
\label{tab:res}
\end{table*}

\subsection{Pole-counting rule}
First we specify in which Riemann sheet one searches for the shadow poles for each $P_c$ state. Taking $P_c(4312)$ as an example, the resonance pole is on the sheet which is unphysical for channels $J/\psi N$, $\bar{D}\Lambda_c$, and $\bar{D}^*\Lambda_c$ but physical for the higher channels $\bar{D}\Sigma_c$, $\bar{D}\Sigma_c^*$, and $\bar{D}^*\Sigma_c$; see Fig.~\ref{fig:thr}. According to the fit results in the previous section, it is a quasi-bound state of $\bar{D}\Sigma_c$ since it lies below and very close to the threshold, and the coupling (residue) to the $\bar{D}\Sigma_c$ channel is large. We expect in the single channel problem $P_c(4312)$ is a physical bound state of $\bar{D}\Sigma_c$, and the shadow pole should be a virtual state on the unphysical sheet of $\bar{D}\Sigma_c$. Hence, in the coupled-channel problem, the corresponding shadow pole should be on the sheet that $J/\psi N$, $\bar{D}\Lambda_c$, $\bar{D}^*\Lambda_c$, and $\bar{D}\Sigma_c$ are all unphysical, while $\bar{D}\Sigma_c^*$ and $\bar{D}^*\Sigma_c$ are still physical. Similar statements hold for the other three states. 

We search for the shadow poles in the complex energy region defined as follows: the imaginary part $-50\,{\rm MeV}<{\rm Im}z<0$ and the real part $\sigma_{n-1}<{\rm Re}z<\sigma_n$, with $\sigma_n$ being the threshold of the main channel and $\sigma_{n-1}$ the lower neighbor channel. For instance, for $P_c(4312)$ (the first pentagram on the very left in Fig.~\ref{fig:thr}), $\sigma_{n}=\sigma_{\bar{D}\Sigma_c}$ and $\sigma_{n-1}=\sigma_{\bar{D}^*\Lambda_c}$. Besides, we search for the shadow poles in the ``elastic'' amplitudes $T_{n\to n}$. 

\begin{figure}[t!]
	\centering
	\begin{subfigure}[b]{0.48\textwidth}
		\centering
		\includegraphics[width=0.8\textwidth]{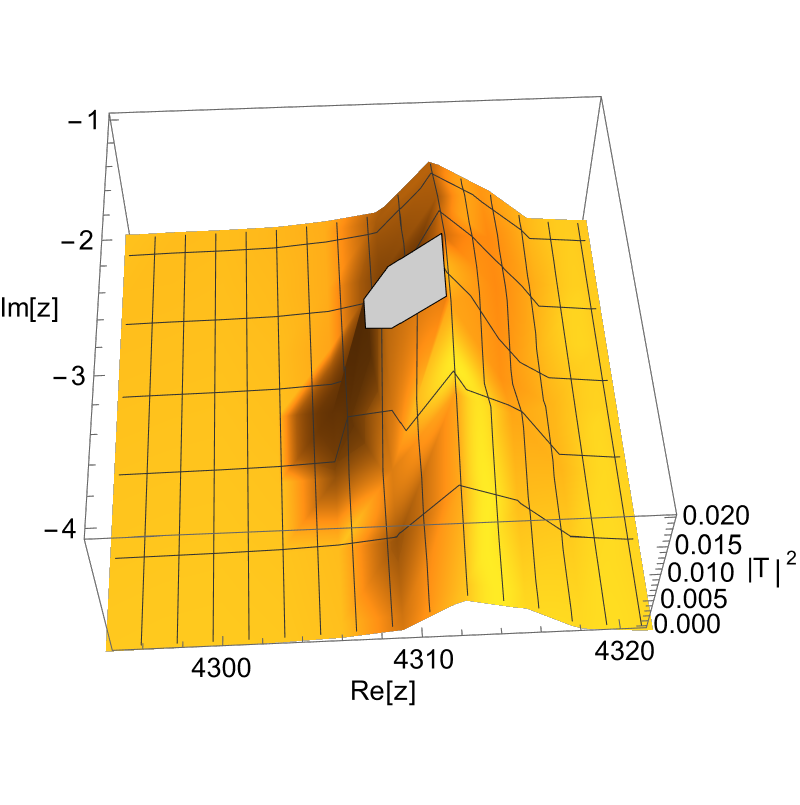}
		\caption{The physical sheet of $\bar{D}\Sigma_c$. }
		\label{fig:4312res}
	\end{subfigure}
	\begin{subfigure}[b]{0.48\textwidth}
		\centering
		\includegraphics[width=0.8\textwidth]{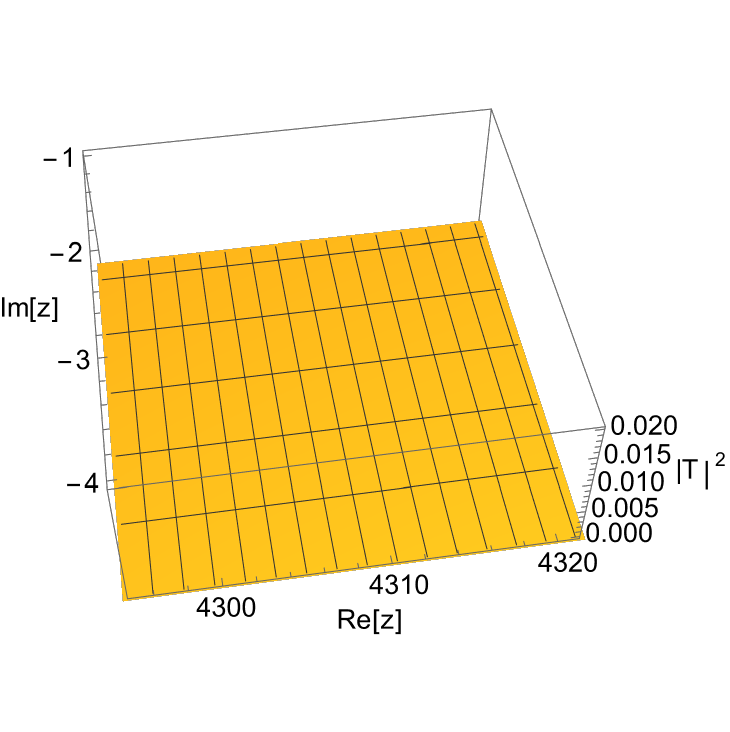}
		\caption{The unphysical sheet of $\bar{D}\Sigma_c$. }
		\label{fig:4312shadow}
	\end{subfigure}
	\caption{The modulus square of the amplitude $T(\bar{D}\Sigma_c\to \bar{D}\Sigma_c)$ with complex energy $z$ (in units of MeV). The $|T|^2$ is in arbitrary units. }\label{fig:4312PC}
\end{figure}

We have not found any shadow poles in the region of concern, for all four $P_c$ states in all three fit solutions. For example, see Fig.~\ref{fig:4312PC}, in fit solution~A in Sec.~\ref{sec:refit}, $P_c(4312)$ is located on the physical sheet of $\bar{D}\Sigma_c$ at $z_p=4313.4-2.2i$ MeV, manifesting itself as a sharp peak in Fig.~\ref{fig:4312res}. However, Fig.~\ref{fig:4312shadow} depicts the unphysical sheet, which contains no structures. 

As a side check, we present the ERE parameters, see Eq.~\eqref{Sera}, obtained from the refined fits (see Table \ref{tab:ERE}). The scattering lengths are all of the order of $1$~fm, with positive real parts, showing a typical feature of amplitudes with bound states. The large imaginary part of the $\bar{D}^*\Sigma_c(c)$ scattering length indicates significant coupled-channel effects, whereas in the other channels, the imaginary parts are rather small. Meanwhile, the effective ranges may contain notable imaginary parts. At a qualitative level, the molecular picture is confirmed since there are no extremely large effective ranges featuring the mechanism of CDD poles. Note that it is hard to make comparisons between our results and the ERE in Ref.~\cite{Guo:2019kdc}, since we have a dynamical coupled-channel approach, providing complex-valued $a$ and $r$. Actually, in Ref.~\cite{Guo:2019kdc}, the pole positions are produced merely by one single channel, resulting in real and negative scattering lengths.
\begin{table}[t!]
    \footnotesize
    \begin{ruledtabular}
    \begin{tabular}{cccc}
    States & Channels & $a$ & $r$\\
    \hline
    $P_c(4312)\, 1/2^-$ & $\bar{D}\Sigma_c$ & \St{$2.0-0.39i$\\$2.2-0.29i$\\$1.9-0.40i$} & \St{$-3.0+2.9i$\\$-3.0+2.9i$\\$-3.0+2.9i$}\\ 
    \hline
    $P_c(4440)\, 1/2^-$ & $\bar{D}^*\Sigma_c(a)$ & \St{$1.2-0.14i$\\$1.3-0.099i$\\$1.3-0.17i$} & \St{$0.36-1.9i$\\$0.44-1.8i$\\$0.43-1.9i$}\\ 
    \hline
    $P_c(4380)\, 3/2^-$ & $\bar{D}\Sigma_c^*(c)$ & \St{$1.9-0.29i$\\$2.0-0.26i$\\$1.9-0.27i$} & \St{$-0.84-1.4i$\\$-0.82-1.3i$\\$-0.83-1.4i$}\\ 
    \hline
    $P_c(4457)\, 3/2^-$ & $\bar{D}^*\Sigma_c(c)$ & \St{$1.2-0.96i$\\$1.2-1.1i$\\$1.1-1.2i$} & \St{$0.78-1.6i$\\$0.77-1.6i$\\$0.76-1.7i$}\\ 
    \end{tabular}
    \end{ruledtabular}
    \caption{The ERE parameters of the main channels for each $P_c$ state (in units of fm), see Eq.~\eqref{Sera}. The notations of the subchannels (a), (b), and (c) are the same as in Table \ref{tab:SWch}. In each cell, the three values from the top to the bottom represent the results in the refined fits A, B, and C, respectively.}
    \label{tab:ERE}
\end{table}
\subsection{Spectral density functions}
\begin{figure*}[t]
	\centering
	\begin{subfigure}[b]{0.48\textwidth}
		\centering
		\includegraphics[width=0.9\textwidth]{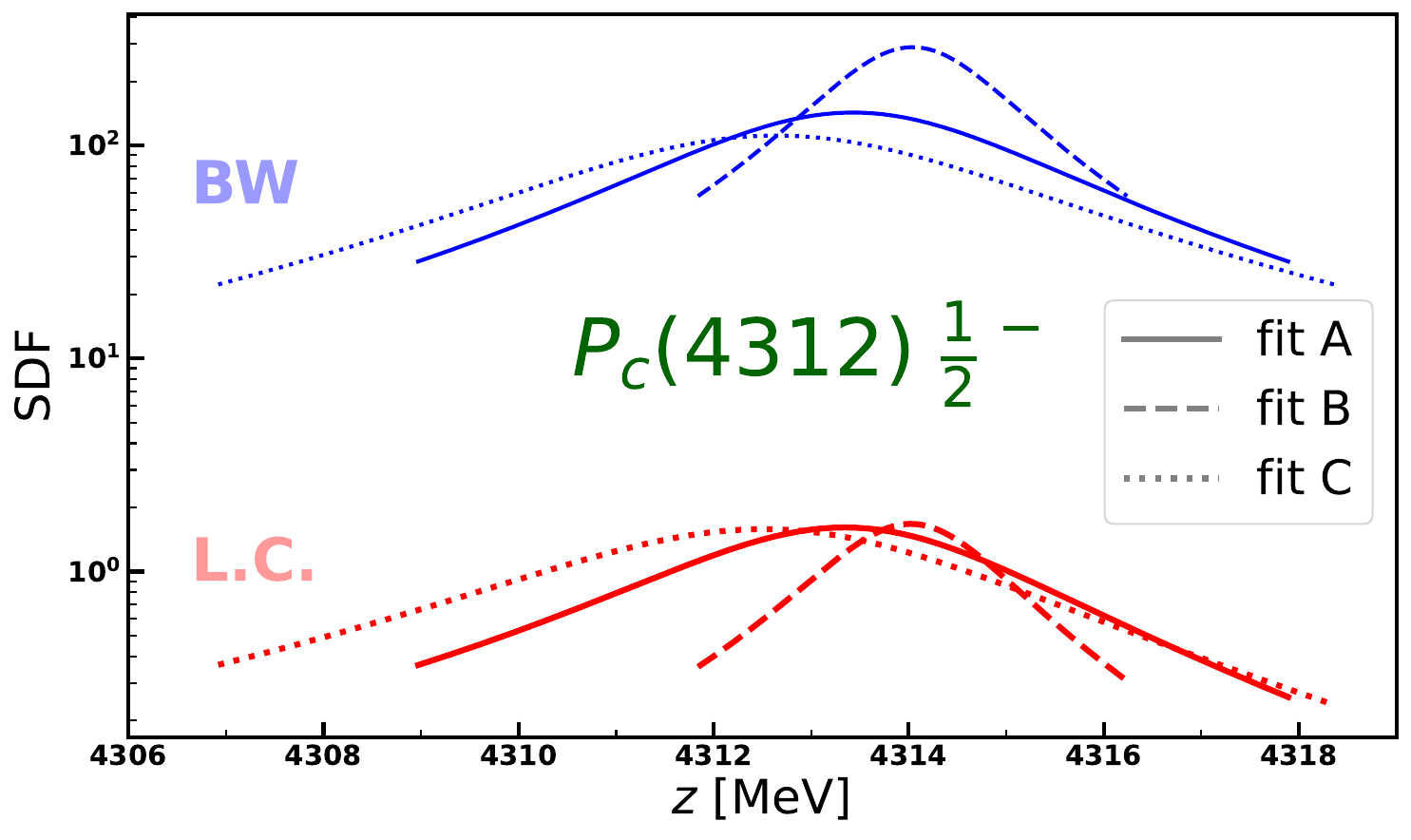}
%		\caption{$P_c(4312)$}
		\label{fig:SDF4312}
	\end{subfigure}
	\begin{subfigure}[b]{0.48\textwidth}
		\centering
		\includegraphics[width=0.9\textwidth]{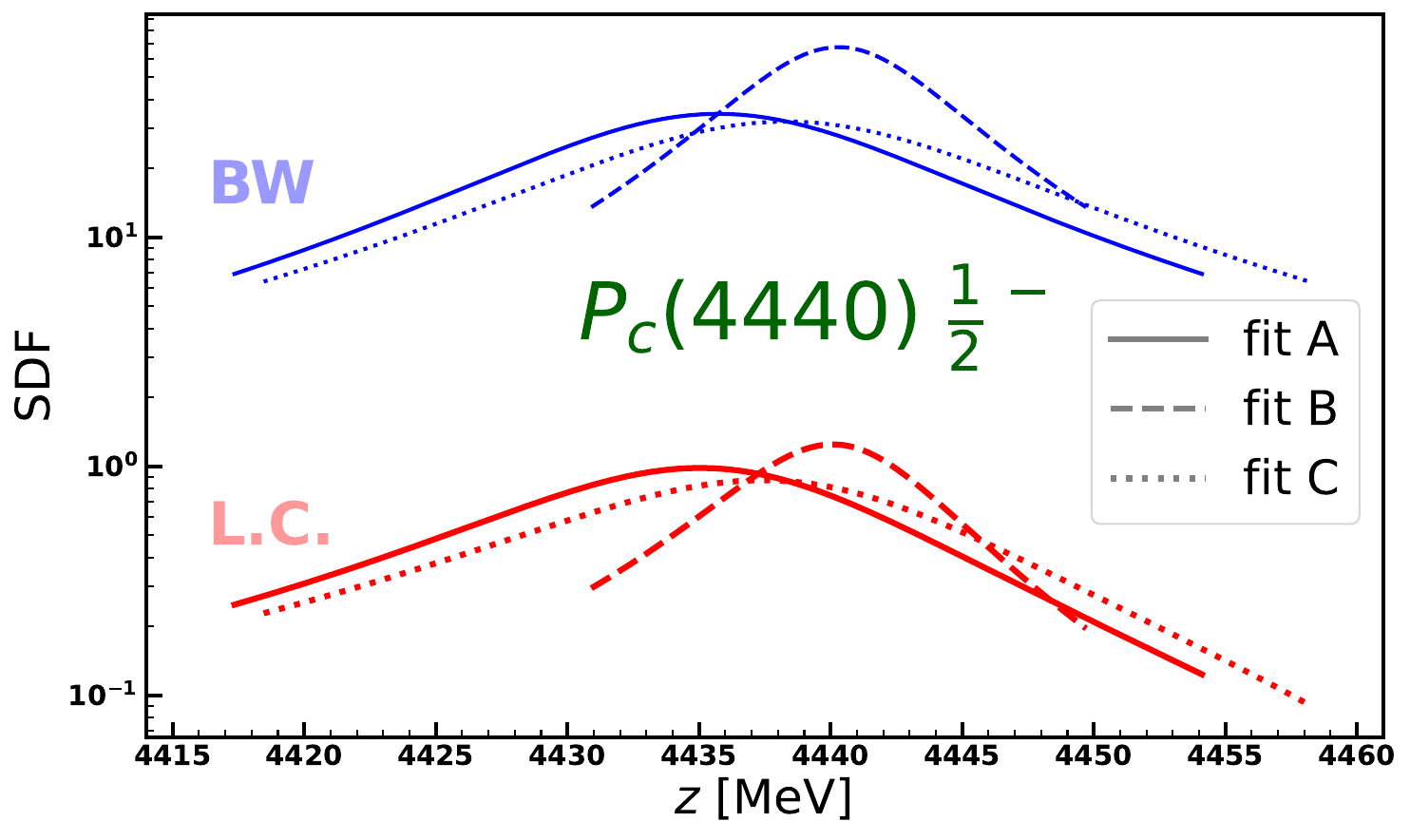}
%		\caption{$P_c(4440)$}
		\label{fig:SDF4440}
	\end{subfigure}
    \begin{subfigure}[b]{0.48\textwidth}
		\centering
		\includegraphics[width=0.9\textwidth]{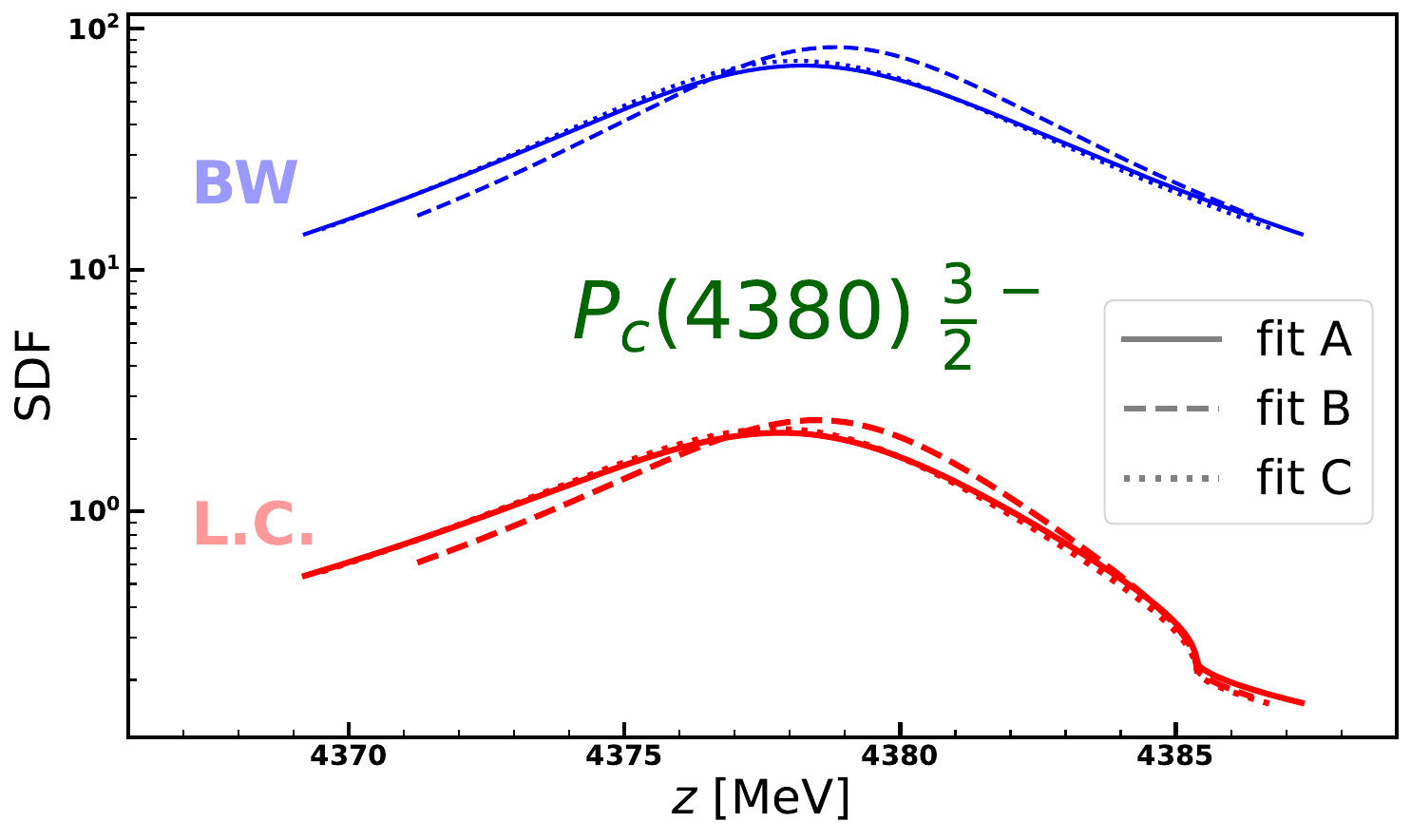}
%		\caption{$P_c(4380)$}
		\label{fig:SDF4380}
	\end{subfigure}
	\begin{subfigure}[b]{0.48\textwidth}
		\centering
		\includegraphics[width=0.9\textwidth]{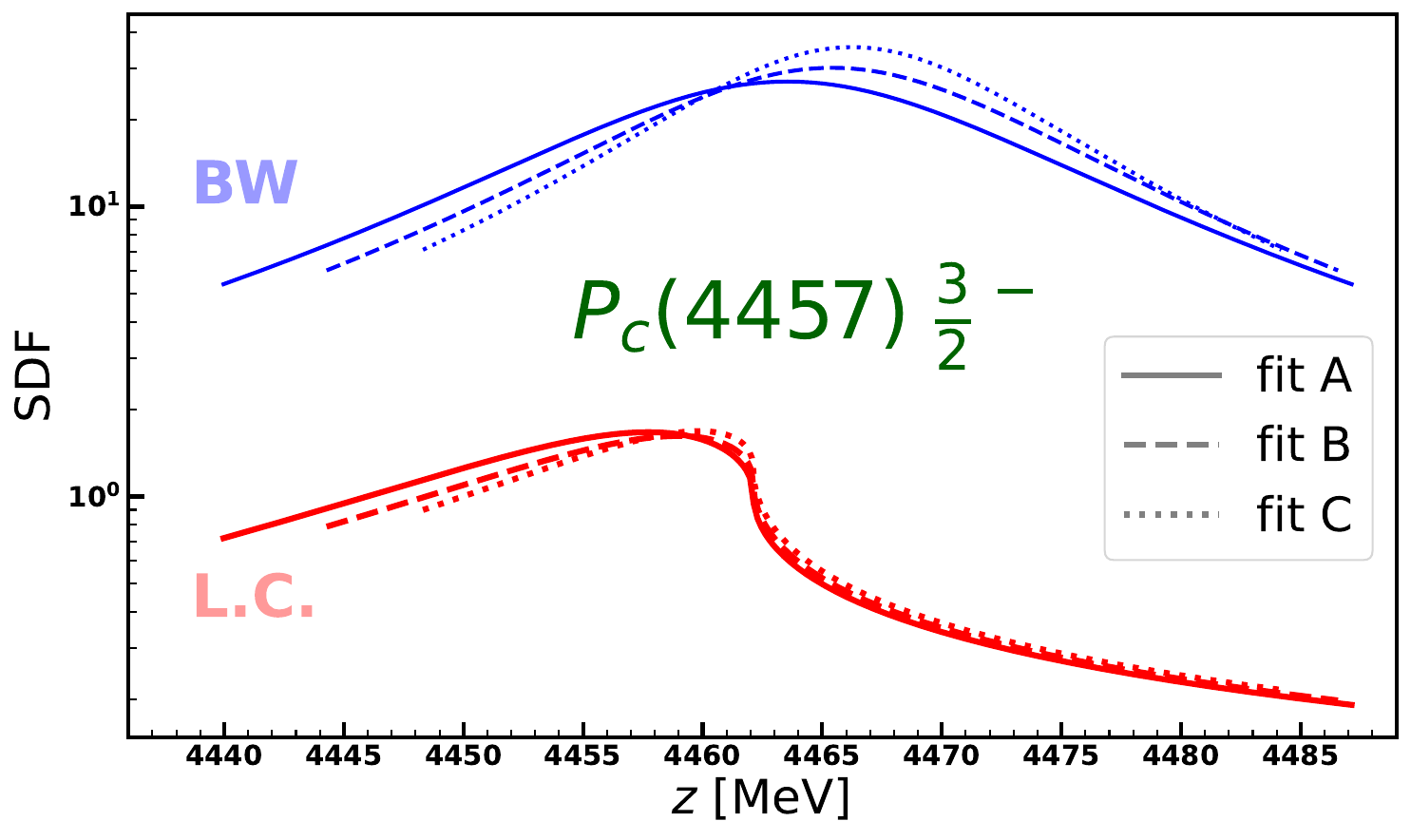}
%		\caption{$P_c(4457)$}
		\label{fig:SDF4457}
	\end{subfigure}
	\caption{The SDFs constructed for the four $P_c$ states in the three fit solutions. In each subfigure, the three lines on the top are the Breit-Wigner (BW) SDFs, while the three on the bottom are the locally constructed (L.C.) SDFs for each pole. Note that the $y$-axis features a logarithmic scale. }\label{fig:SDFplot}
\end{figure*}

Based on the pole positions and residues in Sec.~\ref{sec:refit}, the SDFs defined in Eqs.~\eqref{SDF} and \eqref{NRdef} can be constructed. The elementariness $Z$ is evaluated by Eq.~\eqref{SDFZ}. We plot the SDFs together with the corresponding Breit-Wigner terms in Fig.~\ref{fig:SDFplot}. According to Eq.~\eqref{SDFZ}, the closer the SDF is to the Breit-Wigner term, the bigger the elementariness value gets. Note that the $y$-axis is plotted in logarithmic scale; i.e., in every solution, the constructed SDFs are smaller than the Breit-Wigner terms by one or two orders of magnitude, strongly suggesting the composite picture for all the $P_c$ states in this model. The elementariness values evaluated from Eq.~\eqref{SDFZ} ($Z_{\rm SDF}$), which are almost negligible, are listed in the first row of Table \ref{tab:Zlist}. Additionally, when calculating the $Z_{\rm SDF}$ for the $P_c(4380)$ and $P_c(4457)$ states, the thresholds of $\bar{D}\Sigma_c^*$ and $\bar{D}^*\Sigma_c$ are inside the interval $[M-\Gamma,M+\Gamma]$, causing extra cusps. We have subtracted the physical contributions from such cusps as explained in Appendix~\ref{app:cusp}. 
\begin{table*}[t!]
%\small
\begin{ruledtabular}
	\begin{tabular}{ccccc} 
		Components & $P_c(4312)\, 1/2^-$ & $P_c(4440)\, 1/2^-$ & $P_c(4380)\, 3/2^-$ & $P_c(4457)\, 3/2^-$ \\
        \hline
		$Z_{\rm SDF}$ & \St{$1.1\%$\\$0.6\%$\\$1.4\%$} & \St{$2.8\%$\\$1.8\%$\\$2.7\%$} & \St{$2.9\%$\\$2.7\%$\\$2.8\%$} & \St{$5.1\%$\\$4.4\%$\\$3.8\%$}\\
        \hline
		$Z_{\rm max}$ & \St{$-0.06+0.15i$ ($12.7\%$)\\$0.00+0.14i$ ($11.8\%$)\\$-0.02+0.17i$ ($13.9\%$)} & \St{$0.07+0.28i$ ($22.4\%$)\\$0.09+0.26i$ ($22.2\%$)\\$-0.13+0.44i$ ($27.0\%$)} & 
        \St{$-0.05+0.16i$ ($13.5\%$)\\$-0.03+0.17i$ ($13.9\%$)\\$-0.05+0.15i$ ($13.0\%$)} & \St{$-0.23+0.32i$ ($22.1\%$)\\$-0.23+0.30i$ ($21.6\%$)\\$-0.22+0.23i$ ($19.5\%$)}\\
		\hline
		$X_{\bar{D}\Sigma_c}$ & \St{$1.00-0.14i$ ($82.2\%$)\\$0.94-0.12i$ ($82.2\%$)\\$0.97-0.17i$ ($80.7\%$)} & \St{$-0.00+0.01i$ ($0.4\%$)\\$0.00-0.01i$ ($0.5\%$)\\$-0.00+0.01i$ ($0.5\%$)} & 
        Non $S$-wave & Non $S$-wave\\ 
        \hline
		$X_{\bar{D}^*\Lambda_c(a)}$ & \St{$-0.00+0.01i$ ($0.5\%$)\\$0.00+0.01i$ ($0.6\%$)\\$-0.01+0.01i$ ($0.9\%$)} & \St{$0.00+0.00i$ ($0.3\%$)\\$0.00+0.00i$ ($0.1\%$)\\$0.00+0.00i$ ($0.2\%$)} & 
        Non $S$-wave & Non $S$-wave\\
%        \hline
%		$X_{\bar{D}^*\Lambda_c(b)}$ & $0$ & $0$ & Non $S$-wave & Non $S$-wave\\
        \hline
		$X_{\bar{D}^*\Lambda_c(c)}$ & Non $S$-wave & Non $S$-wave & 
        \St{$0.02+0.02i$ ($2.6\%$)\\$0.02+0.01i$ ($1.7\%$)\\$0.02+0.02i$ ($2.5\%$)} & \St{$-0.00-0.00i$ ($0.1\%$)\\$0.00-0.00i$ ($0.1\%$)\\$-0.00-0.00i$ ($0.1\%$)}\\
        \hline
		$X_{\bar{D}^*\Sigma_c(a)}$ & \St{$0.06-0.01i$ ($4.6\%$)\\$0.06-0.02i$ ($5.4\%$)\\$0.05-0.01i$ ($4.5\%$)} & \St{$0.93-0.28i$ ($76.8\%$)\\$0.91-0.25i$ ($77.2\%$)\\$1.13-0.45i$ ($72.3\%$)} & 
        Non $S$-wave & Non $S$-wave\\
%        \hline
%		$X_{\bar{D}^*\Sigma_c(b)}$ & $0$ & $0$ &  Non $S$-wave & Non $S$-wave\\
        \hline
		$X_{\bar{D}^*\Sigma_c(c)}$ & Non $S$-wave & Non $S$-wave & 
        \St{$0.04-0.04i$ ($4.1\%$)\\$0.03-0.03i$ ($3.4\%$)\\$0.03-0.03i$ ($3.5\%$)} & \St{$1.16-0.12i$ ($66.0\%$)\\$1.18-0.13i$ ($68.5\%$)\\$1.18-0.09i$ ($71.3\%$)}\\
%        \hline
%		$X_{\bar{D}\Sigma_c^*(b)}$ & $0$ & $0$ & Non $S$-wave & Non $S$-wave\\
        \hline
		$X_{\bar{D}\Sigma_c^*(c)}$ & Non $S$-wave & Non $S$-wave & 
        \St{$0.99-0.15i$ ($79.8\%$)\\$0.98-0.15i$ ($81.0\%$)\\$1.00-0.14i$ ($80.9\%$)} & \St{$0.07-0.20i$ ($11.8\%$)\\$0.05-0.16i$ ($9.9\%$)\\$0.05-0.14i$ ($9.0\%$)}\\
    \end{tabular}
\end{ruledtabular}
\caption{The elementariness values evaluated by the SDFs and the complex compositeness/elementariness values from the Gamow wave functions, with the naive measures $\tilde{X}$ and $\tilde{Z}_{\rm max}$ of Eq.~\eqref{ZXp} in the brackets. The notations of the subchannels (a), (b), and (c) are the same as in Table \ref{tab:SWch}. In each cell, the three values from the top to the bottom represent the results in the refined fits~A, B, and C, respectively. Note that only the $S$-wave channels listed in Table \ref{tab:SWch} are considered.}
\label{tab:Zlist}
\end{table*}

\subsection{Compositeness of the Gamow states}
According to Eqs.~\eqref{offR}, \eqref{XX}, and \eqref{ZXp}, we have obtained the complex compositeness and elementariness values of the Gamow states from the off-shell residues given by the three refined fits. The results are also listed in Table \ref{tab:Zlist}. Those results show two obvious features. First, the elementariness values of the four states, though larger than the $Z_{\rm SDF}$ values, are still small in all the fit solutions, indicating the molecular picture. Second, the values coincide the ``quasi-bound states'' picture: on the one hand, the real parts of the compositeness values are all close to the case of a physical bound state, i.e., a probability between $0$ and $1$ (only some of the ${\rm Re}X$ values slightly exceed $1$), and on the other hand, in the channel of the dominant components, the imaginary part of $X$ is much smaller than the real part. 

We would like to note that, as discussed in Ref.~\cite{Wang:2023snv}, the SDFs and the Gamow wave functions are fundamentally two different languages. They are connected to each other only by the physical limit; namely, both of them reproduce Weinberg's criterion when the resonance completely becomes a physical bound state. Hence we cannot expect quantitatively a perfect match of them. Moreover, the naive measure Eq.~\eqref{ZXp} is not unique -- the compositeness of a Gamow state, at its core, is complex-valued. The advantage of the Gamow wave function method is the clear exhibition of the relative proportions among different configurations: in Table \ref{tab:Zlist}, it is obvious that the $\bar{D}\Sigma_c$ configuration dominates the $P_c(4312)$ state, $\bar{D}^*\Sigma_c(a)$ dominates $P_c(4440)$, $\bar{D}\Sigma_c^*(c)$ dominates $P_c(4380)$, and $\bar{D}^*\Sigma_c(c)$ dominates $P_c(4457)$. To this extent, the coupled-channel effects are more significant for the $P_c(4457)$ state, since besides the dominate $\bar{D}^*\Sigma_c(c)$ channel the $X_{\bar{D}\Sigma_c^*(c)}$ proportion is relatively large. This also aligns with the large imaginary part of the $\bar{D}^*\Sigma_c(c)$ scattering length in Table \ref{tab:ERE}. 

We emphasize again that, like many other studies, the $P_c$ states here are all dynamically generated, since the interaction potential $V$ only contains $t$- and $u$-channel exchange diagrams. However, as already mentioned, dynamical generation does not always mean small elementariness. Actually, in the previous study~\cite{Wang:2023snv}, the $N^*(1440)$ state is dynamically generated without a corresponding $s$-channel genuine state, but its elementariness is not negligible. In this study, the conclusion that the $P_c$ states are composite is drawn from careful analyses based on three compositeness criteria that are independent of the model parametrization, rather than simply the fact of dynamical generation. 

\section{Conclusion}\label{sec:con}

In this paper, we investigate the nature of the $P_c$ states based on a dynamical coupled-channel approach. Without any preconceived assumptions as to whether they are hadronic molecules or not, we employ three compositeness criteria: the pole-counting rule, the spectral density functions, and the Gamow wave functions. Those criteria do not depend on the technical details of the model (e.g., whether the poles are induced by $s$-channel genuine states or not), but only on the pole structures and pole parameters -- masses, widths, and couplings (residues) to the scattering channels, which can be extracted from the data by fits. Gathering the results from the criteria and considering the uncertainties given by three fit solutions, there is strong evidence that the four $P_c$ states are all composite: $P_c(4312)\,1/2^-$ is composed by $\bar{D}\Sigma_c$, $P_c(4380)\,3/2^-$ is composed by $\bar{D}\Sigma_c^*$ in the configuration $|J-L|=3/2$, while $P_c(4440)\,1/2^-$ and $P_c(4457)\,3/2^-$ are both recognized as the composite states from $\bar{D}^*\Sigma_c$, in the configurations $S=1/2$ and $S=3/2$, $|J-L|=3/2$, respectively. We have roughly estimated the upper limit of the elementariness of the $P_c$ states, resulting in rather small values. We have also found that the coupled-channel effects are more significant for the $P_c(4457)$, from its relatively larger $\bar{D}\Sigma_c^*$ ($|J-L|=3/2$) proportion and imaginary part of the $\bar{D}^*\Sigma_c$ ($|J-L|=3/2$) scattering length. All in all, these results provide an overall confirmation of the molecular interpretation for the $P_c$ states in the literature. 

An outlook of this paper is pointing at future experiments, for example, the data of photoproductions as proposed in Ref.~\cite{Zhang:2024dkm}. Once the status of the data is significantly improved, not only the $P_c$ spectra but also their structures and properties can be understood more deeply. 

\section*{Acknowledgements}
The authors gratefully acknowledge computing time on the supercomputer JURECA~\cite{JURECA} at Forschungszentrum Jülich under Grant No. {\it baryonspectro}. This work is supported by the National Natural Science Foundation of China under Grant No. 12175240 and the Fundamental Research Funds for the Central Universities. The work of U. G. M. was supported in part by the CAS President's International Fellowship Initiative (PIFI) under Grant No.~2025PD0022. The work of U. G. M. and D. R. is further supported by the MKW NRW under the Funding Code No.~NW21-024-A and by the Deutsche Forschungsgemeinschaft (DFG, German Research Foundation) as part of the CRC 1639 NuMeriQS -- Project No. 511713970.

%\clearpage
\appendix

\section{Subtraction of the cusps in the spectral density functions}\label{app:cusp}
\begin{figure}[t]
	\centering
	\includegraphics[width=0.48\textwidth]{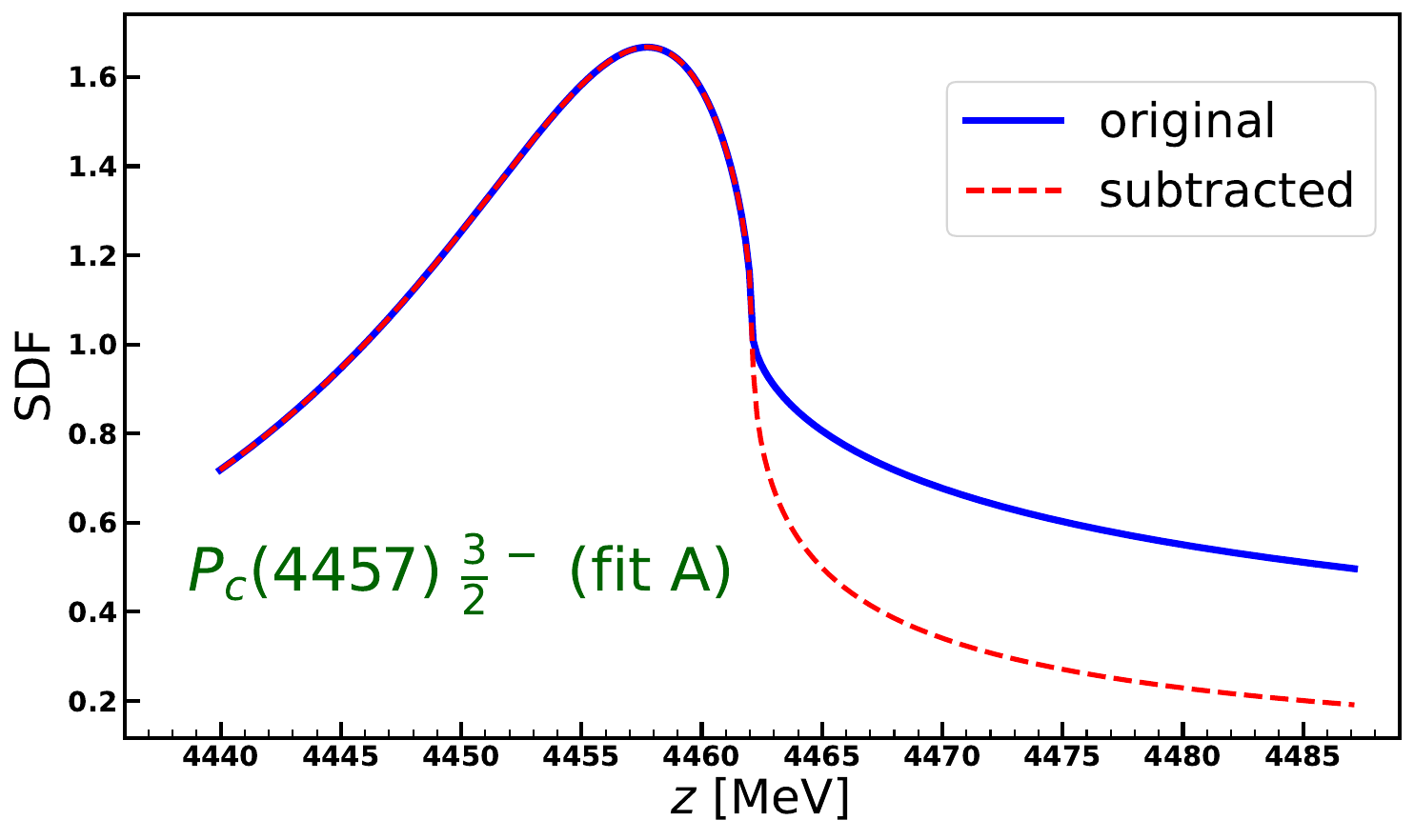}
	\caption{The subtraction of the $\bar{D}^*\Sigma_c$ threshold cusp in the SDF of $P_c(4457)$ (fit A). }
	\label{fig:cuspsub}
\end{figure}

To be concrete, we take the $P_c(4457)$ state in fit~A as an example. According to its pole position in Table \ref{tab:Pcpoles}, the interval $[M-\Gamma,M+\Gamma]$ for the collection of the SDF in Eq.~\eqref{SDFZ} is $[4440,4487]$ MeV, which covers the $\bar{D}^*\Sigma_c$ threshold at $4462$ MeV. For this state, the SDF is
\begin{equation}
    w(z)=
\begin{cases}
    \frac{1}{2\pi}\frac{2h_0p_0+g_1^2 p_1+g_2^2 p_2}{\big|z-M_0+\sum_{\nu=1}^{3} \frac{i}{2} g_\nu^2 p_\nu+ih_0p_0\big|^2} & \quad z\leq \sigma_3\\
    \frac{1}{2\pi}\frac{2h_0p_0+g_1^2 p_1+g_2^2 p_2+g_3^2 p_3}{\big|z-M_0+\sum_{\nu=1}^{3} \frac{i}{2} g_\nu^2 p_\nu+ih_0p_0\big|^2} & \quad z>\sigma_3\\
\end{cases}\ ,
\end{equation}
where the indices $1$, $2$, and $3$ refer to the channels $\bar{D}^*\Lambda_c(c)$, $\bar{D}\Sigma_c^*(c)$, and $\bar{D}^*\Sigma_c(c)$, respectively. This original definition of the SDF has a significant cusp structure due to the sudden appearance of the term $g_3^2 p_3/(\cdots)$, see the solid curve in Fig.~\ref{fig:cuspsub}. The subtraction, as suggested for instance in Ref.~\cite{Kalashnikova:2009gt}, is quite simple -- even if when $z>\sigma_3$, we still exclude the term with $g_3^2 p_3$ in the nominator, 
\begin{equation}\label{SDFsub}
    w_{\rm sub}(z)=\frac{1}{2\pi}\frac{2h_0p_0+g_1^2 p_1+g_2^2 p_2}{\big|z-M_0+\sum_{\nu=1}^{3} \frac{i}{2} g_\nu^2 p_\nu+ih_0p_0\big|^2}\ ,
\end{equation}
then the curve does not contain the new contribution from the third channel; see the dashed line in Fig.~\ref{fig:cuspsub}.

This subtraction is adopted for the $P_c(4380)$ and $P_c(4457)$ resonances in all three refined fit solutions, reducing slightly the estimated $Z_{\rm SDF}$. Nevertheless, even if the SDFs are not subtracted, the $Z_{\rm SDF}$ values remain very small. In addition, the cusp still exists mathematically after the subtraction, since the denominator of Eq.~\eqref{SDFsub} also contains a term $\sim i g_3 p_3/2$. Actually, the subtraction here is a physical operation to exclude the extra contribution from the higher channel.  

\bigskip

\section{Phases of the residues}\label{app:phase}
Consider a two-channel amplitude at a pole $z=z_p$: 
\begin{equation}
    T_{\mu\nu}=\frac{R_{\mu\nu}}{z-z_p}+\cdots\ ,\quad \mu,\nu=1,2\ .
\end{equation}
A fundamental requirement for the matrix of the residues is the ``factorization condition'': 
\begin{equation}
    R_{11}R_{22}=R_{21}R_{12}\ .
\end{equation}
Then it is natural to decompose the residues to the ``couplings'': 
\begin{equation}\label{decomR}
    R_{\mu\nu}\equiv r_\mu r_\nu\ .
\end{equation}
Usually, the condition $R_{12}=R_{21}$ holds and the decomposition is unique once the overall sign of square roots is fixed, for example, the two definitions $r_2^{(A)}\equiv\sqrt{R_{22}}$ and $r_2^{(B)}\equiv R_{12}/\sqrt{R_{11}}$ lead to the same result. However, the time reversal symmetry allows a phase between $R_{12}$ and $R_{21}$, e.g., $R_{12}=-R_{21}$ and $R_{11}R_{22}=-R_{12}^2$; then, 
\[r_2^{(B)}=\frac{R_{12}}{\sqrt{R_{11}}}=\frac{\sqrt{-R_{11}R_{22}}}{\sqrt{R_{11}}}=\pm i\sqrt{R_{22}}\neq r_2^{(A)}\ .\]

Since the unique decomposition~\eqref{decomR} is crucial for the definition of the Gamow wave function, when $R_{12}=-R_{21}$ happens, the phases of the initial and final states should be adjusted, so that in the new matrix of residues $\tilde{R}_{12}=\tilde{R}_{21}$. This has already been mentioned in Sec.~20-d of Ref.~\cite{Taylor:1972pty}. Here, we use the method
\begin{equation}
\begin{split}
    \tilde{\mathcal{R}}&=\begin{pmatrix}e^{i\phi_1}|1\rangle \\ e^{i\phi_2}|2\rangle\end{pmatrix}\hat{R}\big(\langle 1|e^{-i\phi_1},\langle 2|e^{-i\phi_2}\big)\\
    &=
	\begin{pmatrix}
		R_{11} & e^{i(\phi_1-\phi_2)}R_{12} \\
		e^{i(\phi_2-\phi_1)}R_{21} & R_{22} 
	\end{pmatrix}\ ,
\end{split}
\end{equation}
where $R_{\mu\nu}\equiv\langle\mu|\hat{R}|\nu\rangle$. Setting $\phi_1=0$ and $\phi_2=\pi/2$, we have the new elements $\tilde{R}_{12}=-iR_{12}$ and $\tilde{R}_{21}=+iR_{21}$, satisfying $\tilde{R}_{12}=\tilde{R}_{21}$. 

Actually, in the current model, the time-reversal phase occurs $R_{\alpha V}=-R_{V \alpha}$ if $V$ is a channel containing $\bar{D}^*$ while $\alpha$ is not. For every such kind of element, we multiply the phase $\pm i$. Note that this phase redefinition is related to neither the dynamics nor to the experimental observables; it only maintains the mathematical consistence when applying the Gamow wave function method. 

\bibliographystyle{h-physrev}
\bibliography{Pc2.bib}

\end{document}